\documentclass[conference]{IEEEtran}
\IEEEoverridecommandlockouts

\def\BibTeX{{\rm B\kern-.05em{\sc i\kern-.025em b}\kern-.08em
    T\kern-.1667em\lower.7ex\hbox{E}\kern-.125emX}}
    
\usepackage{subcaption}
\usepackage{balance}
\usepackage[inline]{enumitem}
\usepackage{booktabs}
\usepackage[acronym]{glossaries}
\usepackage{multirow}
\usepackage{url}
\usepackage{amssymb}
\usepackage{pifont}
\newcommand{\cmark}{\ding{51}}
\newcommand{\xmark}{\ding{55}}

\usepackage{tikz}

\newacronym{ACT}{ACT}{Automatic Contact Tracing}
\newacronym{BLE}{BLE}{Bluetooth Low-Energy}
\newacronym{PEPP-PT}{PEPP-PT}{Pan-European Privacy-Preserving Proximity Tracing}
\newacronym{ROBERT}{ROBERT}{ROBust and privacy-presERving proximity Tracing}
\newacronym{DP-3T}{DP-3T}{Decentralized Privacy-Preserving Proximity Tracing}
\newacronym{OS}{OS}{Operating System}
\newacronym{API}{API}{Application Programming Interface}
\newacronym{GAEN}{GAEN}{Google Apple Exposure Notification}
\newacronym{DoS}{DoS}{Denial-of-Service}
\newacronym{L2CAP}{L2CAP}{Logical Link Control and Adaptation Protocol}
\newacronym{RFCOMM}{RFCOMM}{Radio Frequency Communication}
\newacronym{SDP}{SDP}{Session Description Protocol}
\newacronym{ATT}{ATT}{Attribute Protocol}
\newacronym{GATT}{GATT}{Generic Attribute Profile}
\newacronym{HCI}{HCI}{Host Controller Interface}
\newacronym{ID}{ID}{IDentifier}
\newacronym{EBID}{EBID}{Ephemeral Bluetooth Identifier}
\newacronym{SK}{SK}{Secret Key}
\newacronym{TEK}{TEK}{Temporary Exposure Key}
\newacronym{RPI}{RPI}{Rotating Proximity Identifier}
\newacronym{AEM}{AEM}{Associated Encrypted Metadata}
\newacronym{UserID}{UserID}{User IDentifier}
\newacronym{TempID}{TempID}{Temporary IDentifier}
\newacronym{RPIK}{RPIK}{Rolling Proximity Identifier Key}
\newacronym{AEMK}{AEMK}{Associated Encrypted Metadata Key}
\newacronym{OTP}{OTP}{One Time Password}

\makeglossaries{}  
    
\begin{document}

\title{Contact Tracing Made Un-\emph{relay}-able\\}

\author{\IEEEauthorblockN{Marco Casagrande}
\IEEEauthorblockA{\textit{Department of Mathematics} \\
\textit{University of Padua}\\
Padua, Italy \\
 m.casagrande.1993@gmail.com}
\and
\IEEEauthorblockN{Mauro Conti}
\IEEEauthorblockA{\textit{Department of Mathematics} \\
\textit{University of Padua}\\
Padua, Italy \\
conti@math.unipd.it}
\and
\IEEEauthorblockN{Eleonora Losiouk}
\IEEEauthorblockA{\textit{Department of Mathematics} \\
\textit{University of Padua}\\
Padua, Italy \\
eleonora.losiouk@unipd.it}
}

\maketitle

\begin{abstract}
Automated contact tracing is a key solution to control the spread of airborne transmittable diseases: it traces contacts among individuals in order to alert people about their potential risk of being infected. The current SARS-CoV-2 pandemic put a heavy strain on the healthcare system of many countries. Governments chose different approaches to face the spread of the virus and the contact tracing apps were considered the most effective ones. In particular, by leveraging on the Bluetooth Low-Energy technology, mobile apps allow to achieve a privacy-preserving contact tracing of citizens. While researchers proposed several contact tracing approaches, each government developed its own national contact tracing app.

In this paper, we demonstrate that many popular contact tracing apps (e.g., the ones promoted by the Italian, French, Swiss government) are vulnerable to relay attacks. Through such attacks people might get misleadingly diagnosed as positive to SARS-CoV-2, thus being enforced to quarantine and eventually leading to a breakdown of the healthcare system. To tackle this vulnerability, we propose a novel and lightweight solution that prevents relay attacks, while providing the same privacy-preserving features as the current approaches. To evaluate the feasibility of both the relay attack and our novel defence mechanism, we developed a proof of concept against the Italian contact tracing app (i.e., \emph{Immuni}). The design of our defence allows it to be integrated into any contact tracing app. 
\end{abstract}


\section{Introduction}
\label{sec:introduction}

The SARS-CoV-2 pandemic caught governments and healthcare systems unprepared. One of the main issues of the virus concerns the speed of diffusion, which is very high, especially in comparison to the time required to find all the people that have been in contact with an infected person. Thus, the very first strategy adopted by some governments was to put citizens in a strict lockdown. However, the lockdown was only a temporary solution and governments started looking for alternative approaches aimed at the containment of the virus, which means: 
\begin{enumerate*}[label=(\roman*)]
    \item rapid identification of the infected individuals, with the consequent quarantine,
    \item identification of the people that have been close to an infected person in the previous days and weeks,
    \item decontamination of places where the infected individual has been.
\end{enumerate*}
Guaranteeing the containment is a difficult task, which might lead to errors, especially if performed manually, as it happened at the start of the virus spread . As a result, governments expressed their need for \gls{ACT} solutions, which can help with better monitoring the spread of the virus if applied together with social distancing. 

Several \gls{ACT} solutions were proposed and adopted by different countries according to their economical, technological and cultural status. The most promising tool for collecting contact tracing data that concern the citizens was the smartphone, since most people own one. Tracing citizens contacts through the smartphone location (i.e., GPS signal, Wi-Fi routers, cellular networks) seemed a feasible solution~\cite{crepuscolo}, which, however, introduced significant privacy issues: citizens had to partially give up on their privacy to protect others. 
As illustrated in~\cite{rel-rogue}, such privacy issues could even lead to public identification of diagnosed patients or to mass surveillance. Thus, researchers proposed an alternative approach based on the \gls{BLE} technology, which is already available on most smartphones. BLE allows two devices that are close to each other to exchange their identifiers, while requiring a limited amount of battery power. Several communication protocols have been proposed to support ACT on mobile devices: BlueTrace~\cite{web-bluetrace}, \gls{ROBERT}~\cite{web-robert} and  \gls{DP-3T}~\cite{web-dp3t}. In addition, Google and Apple designed the \gls{GAEN}~\cite{web-gaen} protocol. \gls{GAEN} enables the interoperability among devices running Android and iOS, by providing a common API in the underlying platform to be used by contact tracing apps. All the above-mentioned protocols follow the same workflow: each smartphone has its own pseudonym, shared with other smartphones when they get close to each other. After a while, the smartphone updates its pseudonym with a new one, which is seemingly independent. Thus, each smartphone will soon have a database of the announced pseudonyms and a database of the received pseudonyms. When a person is found infected, all the smartphones that have been in contact with the smartphone of the infected person should be notified, also receiving a risk score. By proving the unlinkability of pseudonyms, the above approach guarantees a good level of privacy. Despite its privacy-preserving nature, the BLE-based \gls{ACT} approach has several limitations which reduce the citizens consensus to adopt it. Among such limitations, there is the resilience to security attacks. In particular, the absence of an authentication procedure in \gls{ACT} apps paves the way for replay and relay attacks. 

In this paper, we focus on the vulnerability to relay attacks of \gls{ACT} apps and we prove that many popular ones (e.g., those promoted by the Italian, French, Swiss government) are not able to defend against such attacks. To tackle this security issue, we designed a solution called \emph{ACTGuard} that effectively prevents relay attacks by acquiring the location data concerning contacts between pairs of people. In particular, \emph{ACTGuard} saves on the mobile device only a hash of the contact, generated by providing as the input of a hash function the pseudonyms of the two smartphones, the timestamp of the contact and the location of the contact. Whenever a person is found infected, the hashes of the contacts over the previous days are shared with a remote server and locally downloaded by other smartphones. If the pseudonym of the infected person has been spoofed and used in a relay attack (i.e., being re-transmitted by the attacker in a different location than the person real one), the hashes of the relay attack victims will mismatch the ones of the infected person due to the different location. This way, \emph{ACTGuard} will prevent the relay attack victims from receiving a false positive alert. Moreover, by saving and sharing only the hashes of the contacts, \emph{ACTGuard} provides a privacy-preserving solution. To assess the feasibility of the relay attack against \gls{ACT} apps and the reliability of \emph{ACTGuard}, we developed two proofs of concept focusing on \emph{Immuni}, the Italian contact tracing app based on GAEN.

\textbf{Contributions.} The contributions of the paper are as follows: 
\begin{itemize}
    \item We analyze the design of popular \gls{ACT} apps based on \gls{GAEN} and identified a vulnerability against relay attacks;
    \item We design the first solution that is compliant with the current \gls{GAEN} internal architecture and prevents relay attacks, called \emph{ACTGuard};
    \item We implemented a proof of concept of a relay attack and of \emph{ACTGuard} against \emph{Immuni}, the Italian \gls{ACT} app;
    \item We release a video demonstration of the attack and of the defence\footnote{\url{https://spritz.math.unipd.it/projects/immuniguard/demo/immuniguard-demo.avi}}.
\end{itemize}

\textbf{Organization}
The rest of the paper is organized as follows: Section~\ref{sec:background} provides background knowledge about the \gls{BLE} technology, and the existing proximity tracing solutions for \gls{ACT} apps; Section~\ref{sec:system_model_threat_model} introduces the system model and the threat model of the relay attack we designed against \gls{GAEN}-based \gls{ACT} apps; Section~\ref{sec:ACTGuard_design} outlines the design of \emph{ACTGuard}; Section~\ref{sec:immuniguard_design} describes our implementation of the relay attack against \emph{Immuni} and our proof of concept of \emph{ACTGuard}, i.e., \emph{ImmuniGuard}; Section~\ref{sec:related_work} illustrates the related work concerning security and privacy issues of \gls{ACT} apps; finally, in Section~\ref{sec:discussion}, we conclude the paper with a discussion about \gls{ACT} apps. 
\section{Background}
\label{sec:background}
The purpose of this section is to illustrate the \gls{BLE} protocol (i.e., Section~\ref{ssec:bluetooth}) and the proximity tracing protocols (i.e., Section~\ref{ssec:act_frameworks}). The description of the \gls{BLE} protocol focuses on the Android \gls{OS}, since both the attack and the defence target this platform. 

\subsection{Bluetooth}
\label{ssec:bluetooth}

\textbf{Bluetooth stack.} The Bluetooth stack involves four different layers: physical, link, middleware and application layer. The physical layer and the link layer include physical and hardware components, such as chips, that communicate with the \gls{OS} through the \gls{HCI}. The middleware layer includes the protocols implemented by the host, among which the \gls{L2CAP}, the \gls{RFCOMM} and the \gls{SDP}. \gls{L2CAP} is responsible for the Bluetooth data-flow control, \gls{RFCOMM} generates serial data stream, while \gls{SDP} broadcasts the device services to other devices in order to establish a connection. Finally, the application layer defines the different functionalities offered to users. From version 4.0, the Bluetooth specification includes also the \gls{BLE} version, supporting the transmission of discrete data to reduce power consumption. In BLE, discrete data are stored in \emph{attributes}, which access is regulated by the \gls{ATT} and the \gls{GATT} protocols. BLE devices expose their data through \emph{services}, that include \emph{attributes} and \emph{characteristics} (i.e., special types of \emph{attributes}). 
 
\textbf{Bluetooth protocol.} The Bluetooth protocol always involves two devices, i.e., a client and a server. Usually, the server keeps running in discoverable mode until a client sends it an inquiry. As soon as the inquiry is received, the server sends its information to the client (i.e., device name, device class, list of services, technical information, a user-friendly name defined by the manufacturer). Being aware of the server's services, the client attempts to read such data. However, the access is regulated by the set of permissions associated to the data: no permission (the server can access the client data); authentication required (the client starts the \emph{pairing procedure}); authorization required (any implementation concerning the authorization is left to developers). 
The \emph{pairing procedure} allows a device to authenticate a different, remote one, before connecting to it, and to share with it long-term keys that will encrypt the communication. During the authentication, the user is asked to verify the identity of the remote device. This process might go through an interaction with the remote device display or through the input of a value into the remote device, according to its features. Alternatively, the \emph{pairing procedure} might also happen through the \emph{just works} mode, in case the remote device has neither a display nor an input. When the \emph{pairing} is completed, the two devices go through the \emph{bonding procedure}, during which they exchange the long-term keys. Unless a device is manually reset or unpaired, it will keep the key stored and use it to encrypt the communication at the link layer. 

\textbf{Use of Bluetooth in Android.} In Android, applications willing to use the Bluetooth communication channel have to rely on the \texttt{android.bluetooth} API package, that enables the interaction with the system process under \texttt{/packages/apps/Bluetooth}. Moreover, since Bluetooth is classified as a protected resource, applications have to declare specific permissions, according to their purposes: 
\begin{itemize}
    \item \texttt{BLUETOOTH}, to support Bluetooth communications, such as requesting and accepting connections.
    \item \texttt{BLUETOOTH\_ADMIN}, to discover nearby Bluetooth devices and to change smartphone Bluetooth setting. Currently, this permission requires also location related permissions, such as \texttt{android.permission.ACCESS\_FINE\_LOCATION} or \\ \texttt{ACCESS\_COARSE\_LOCATION}.  
    \item \texttt{BLUETOOTH\_PRIVILEGED}, to avoid any user interaction with the device during the \emph{pairing procedure} (it can be granted only to system applications with a signature-level permission). 
\end{itemize}
When an application is granted the required permissions, it can start communicating with external Bluetooth devices. If an external device requires a \emph{pairing procedure}, this is usually handled by a single application. However, the bonding information of the external device is saved in a shared location, that can be accessed by any application with Bluetooth related permissions. As soon as it has the permissions, any application can communicate with an external Bluetooth device, if this has been already paired with the smartphone.  

\subsection{Proximity Tracing Protocols}
\label{ssec:act_frameworks}
\begin{figure}[h!]
	\centering
    \includegraphics[width=0.9\linewidth]{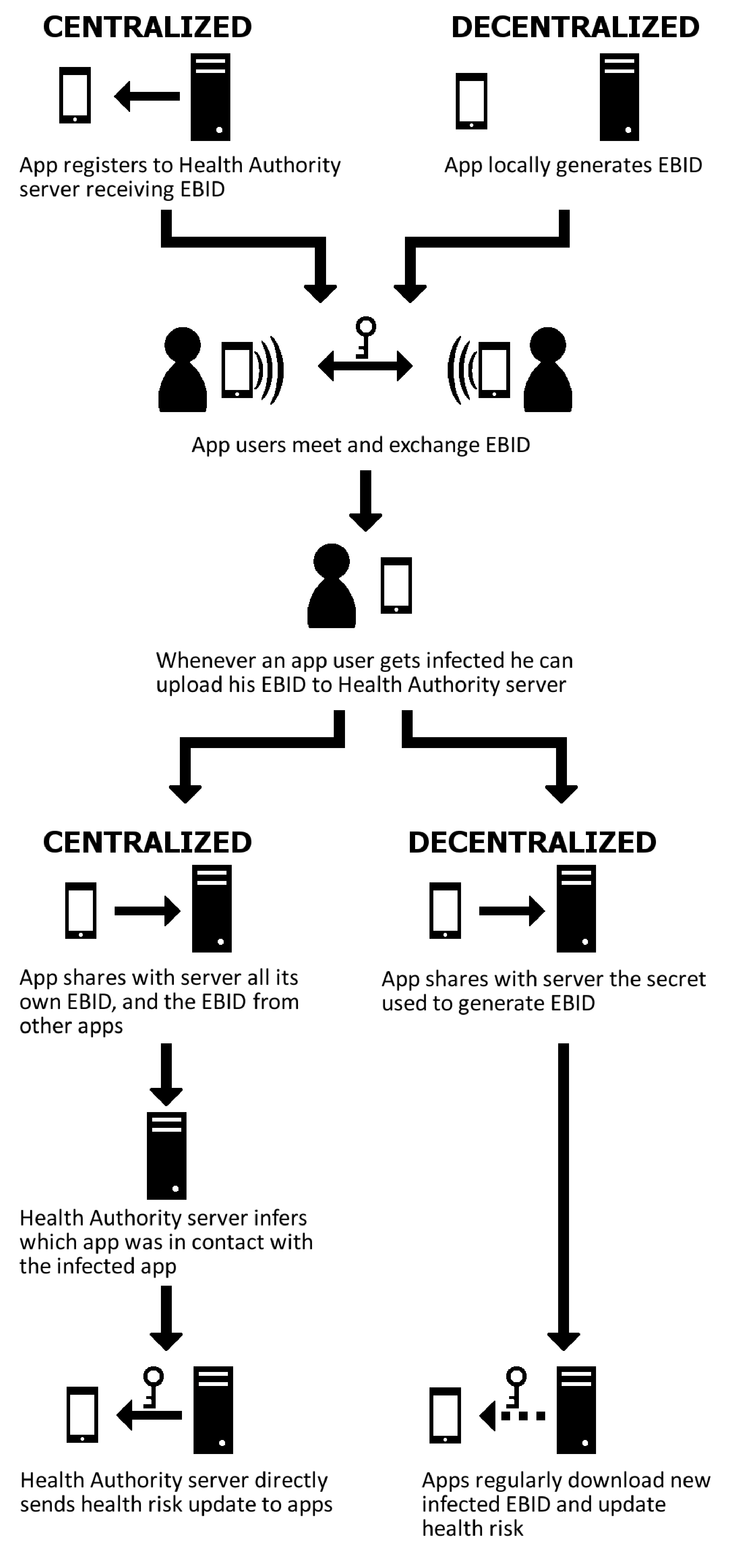}
	\caption{Comparison between centralized and decentralized approaches.}
	\label{fig:centvsdecent}
\end{figure}

BlueTrace~\cite{web-bluetrace}, \gls{ROBERT}~\cite{web-robert},  \gls{DP-3T}~\cite{web-dp3t} and \gls{GAEN}~\cite{web-gaen} are among the most well-known proximity tracing protocols worldwide. Proximity tracing consists in registering any physical contact between individuals, and it is a core part of the contact tracing process. Despite sharing the same technology (i.e., \gls{BLE} on mobile devices), such protocols differ in the role assigned to the centralized server, which is controlled by the national health authority. According to this, the protocols can be classified into two different approaches, both shown in Fig.~\ref{fig:centvsdecent}: \emph{centralized} and \emph{decentralized}. The general workflow of proximity tracing protocols can be summarized as follows: 
\begin{itemize}
    \item Set up and configuration: in the \emph{centralized} approach, the app registers to the health authority server and it receives multiple \gls{EBID}. On the contrary, the \emph{decentralized} approach assumes that the app locally generates its own \gls{EBID}. 
    \item Contact between two individuals: when two app users meet each other, the respective apps share their own \gls{EBID}.  
    \item Announcement of a new positive: when an app user is found infected, the \emph{centralized} approach requires the user to share with the health authority server all his own \gls{EBID} and the ones of the people met. On the contrary, the \emph{decentralized} approach requires the infected user to share only his own \gls{EBID} remotely. 
    \item Risk score calculation: in the \emph{centralized} approach, the health authority server calculates the risk for all the \gls{EBID} that have been in contact with the infected one and it sends the notifications accordingly. In the \emph{decentralized} approach, the apps periodically download the \gls{EBID} of the new infected people and locally calculate the risk score. 
\end{itemize}

We will now provide more details about each proximity tracing protocol.

\textbf{BlueTrace.} BlueTrace~\cite{rel-bluetrace} is a proximity tracing protocol designed by the Government Digital Services team at the Government Technology Agency of Singapore and it adheres to the \emph{centralized} approach. It was implemented on the first national \gls{ACT} app deployed in the world, i.e., TraceTogether~\cite{web-bluetrace}. In this case, the \emph{centralized} approach was chosen for two reasons: 1) it allows for a human-in-the-loop design during epidemiological surveillance, and 2) it allows to better monitor the adoption and usage of the app, by logging daily requests to the server.
During app registration, the centralized server assigns the user a unique \gls{UserID}, linked to his phone number. Then, the app generates multiple \gls{TempID} from the UserID, thus allowing the centralized server to trace back the \gls{UserID} from the \gls{TempID}. The messages exchanged between apps as advertisement packets contain the following data:
\begin{itemize}
    \item \gls{TempID} - they frequently rotate out to prevent third-party tracking;
    \item Device Model - it improves the calculation of distance estimates;
    \item Organization Code - it indicates the country and health authority;
    \item BlueTrace Protocol Version.
\end{itemize}

When a user is found positive to SARS-CoV-2, BlueTrace allows him to share his contacts history with the centralized server, together with additional data (e.g., RSSI, device model, timestamp). The server decrypts the \gls{TempID} in order to retrieve the \gls{UserID} and the validity period. Then, it finds the closest contacts according to time of exposure and distance. During the medical interview of infected users, proximity and duration filtering thresholds are fine-tuned, before any in-app exposure warning is sent. 

\textbf{ROBERT.} PEPP-PT~\cite{web-pepppt} is a non-profit organization, based in Switzerland, with 130 members across eight European countries. The objective of PEPP-PT is to develop a privacy-preserving digital proximity-tracing architecture. Part of the PEPP-PT proposal involved the design of a \emph{centralized} contact tracing protocol, named \gls{ROBERT}~\cite{web-robert}. During app registration, the centralized server assigns to the user a permanent \gls{ID}, saved in its database, and several \gls{EBID}, which are valid within a specific time window. Apps implementing the ROBERT protocol regularly broadcast packets containing their \gls{EBID}, and store \gls{EBID} received from other users. Whenever a user is found positive to SARS-CoV-2, ROBERT requires him to upload the list of \gls{EBID} collected from other users in the past weeks. The centralized server is able to trace back the ID from uploaded \gls{EBID}, flags them as exposed, calculates the risk score and warns app users accordingly. 

\textbf{DP-3T.} DP-3T~\cite{web-dp3t} is a European consortium of technologists, legal experts, engineers and epidemiologists with the primary objective of developing a proximity tracing protocol able to prevent mass surveillance. The DP-3T protocol adheres to the \emph{decentralized} approach and provides three slightly different designs:
\begin{enumerate*}[label=(\roman*)]
    \item a \emph{low-cost decentralized proximity tracing} protocol (good privacy and very small bandwidth required);
    \item an \emph{unlinkable decentralized proximity tracing} protocol (far better privacy than the first solution, but with an increased bandwidth requirement);
    \item a \emph{hybrid decentralized proximity tracing} protocol (a combination of the two previous designs).
\end{enumerate*}
From now on, we will implicitly refer to the \emph{hybrid decentralized proximity tracing} design, as it is more well-rounded: it features a high level of protection against linking EBID to the real identity of individuals, while still retaining a low-cost philosophy. An app implementing DP-3T does not communicate with the centralized server during the setup process. On the contrary, it locally generates a \gls{SK}, used to generate several \gls{EBID}, which are valid within a specific time window. Apps implementing DP-3T regularly broadcast packets containing their \gls{EBID}, and store the \gls{EBID} received from other users. Whenever a user is found positive to SARS-CoV-2, DP-3T requires him to upload past \gls{SK}, along with the associated time windows. The centralized server simply acts as a database, allowing other app users to connect and download the newly uploaded data. By knowing a \gls{SK} and its time windows, apps can generate \gls{EBID} and compare them to the ones they collected during their proximity tracing activity. The health risk calculation is performed locally by the app, and the health status is updated immediately.

\textbf{GAEN.} The result of the collaboration between Google and Apple is the \gls{GAEN}~\cite{web-gaen} protocol, which aims at enabling \gls{BLE} interoperability between Android and iOS devices. \gls{GAEN} provides a set of \gls{API} that are restricted only to the developers authorized by their own Government to release a national \gls{ACT} app. \gls{GAEN} shares several design features with the DP-3T \emph{hybrid decentralized proximity tracing} and it adheres to the \emph{decentralized} approach (for more details on the \gls{GAEN} protocol, please, refer to Section~\ref{sec:system_model_threat_model}).  

\textbf{Overview of \gls{ACT} Apps.} As already mentioned, BlueTrace, \gls{ROBERT}, \gls{DP-3T} and \gls{GAEN} are proximity tracing protocols, which can be implemented by any national \gls{ACT} app. Table~\ref{tab:relworkover} shows an overview of the \gls{ACT} apps that we analyzed. During our selection, we aimed to include entries for each of the most popular proximity tracing protocols, to represent a wide variety of countries and to showcase the various combination of technologies and privacy approaches with respect to the following criteria: the adopted proximity tracing protocol, the technology used for performing the contact tracing, the set of personal data required to be shared by the app user and, finally, the resilience to replay and relay attacks. As shown in the table, almost all apps are vulnerable to relay attacks. This is the main motivation that encouraged us to provide a design for a defence mechanisms, that prevents such attacks, while being compliant to existing protocols (for more details on the \gls{ACT} apps, please, refer to Section~\ref{sec:related_work}).

\begin{table*}[ht!]
	\centering
    \begin{tabular}{|l|l|l|c|c|l|c|c|}
        \hline
        \textbf{\gls{ACT} App} & \textbf{Country} & \textbf{Proximity Tracing Protocol} & \textbf{GPS} & \textbf{Bluetooth} & \textbf{Data Shared} & \textbf{Replay} & \textbf{Relay} \\ \hline
        
        TraceTogether~\cite{web-bluetrace} & Singapore & OpenTrace & \xmark & \xmark & Phone Number & \cmark & \cmark \\
        Safe Paths~\cite{rel-rogue} & United States & PACT & \cmark & \xmark & GPS Location & \cmark & \xmark \\
        Hamagen~\cite{ct-hamagen} & Israel & Proprietary Protocol & \cmark & \xmark & GPS Location & \cmark & \cmark \\
        Aarongya Setu~\cite{ct-aarongyasetu} & India & Proprietary Protocol & \cmark & \xmark & Personal Data & \cmark & \cmark \\
        StopCovid~\cite{ct-stopcovid} & France & ROBERT & \xmark & \xmark & Pseudonyms & \cmark & \cmark \\
        SwissCovid~\cite{ct-swisscovid} & Switzerland & GAEN & \xmark & \cmark & Pseudonyms & \cmark & \cmark \\
        Covid Alert NY~\cite{ct-alertny} & United States & GAEN & \xmark & \cmark & Pseudonyms & \cmark & \cmark \\
        Immuni~\cite{ct-immuni} & Italy & GAEN & \xmark & \cmark & Pseudonyms & \cmark & \cmark \\
        \hline
    \end{tabular}
    \vspace*{1mm}
    \caption{Overview of the Analyzed \gls{ACT} Apps.}
    \label{tab:relworkover}
\end{table*}

\section{System Model and Threat Model}
\label{sec:system_model_threat_model}
In this section, we illustrate the system model (i.e., Section~\ref{ssec:system_model}) and the threat model (i.e., Section~\ref{ssec:threat_model}) we considered for designing a relay attack against \gls{ACT} apps and for our defence mechanism, i.e., \emph{ACTGuard}. In particular, among the different proximity tracing protocols, we chose to focus on \gls{GAEN}-based contact tracing apps, as \gls{GAEN} is the most widely adopted solution in Europe.

\subsection{System Model}
\label{ssec:system_model}

Most \gls{GAEN}-based apps merely act as an interface, that strongly relies on the underlying protocol. Thus, our security analysis concerning the resilience to relay attack, as well as the design of \emph{ACTGuard}, can be applied to any GAEN-based contact tracing app. Here, we provide the details of the \gls{GAEN} protocol in terms of pseudonym generation and exchange, as well as announcement of a new positive person.  

\begin{figure}[h!]
  \centering
  \includegraphics[width=\linewidth]{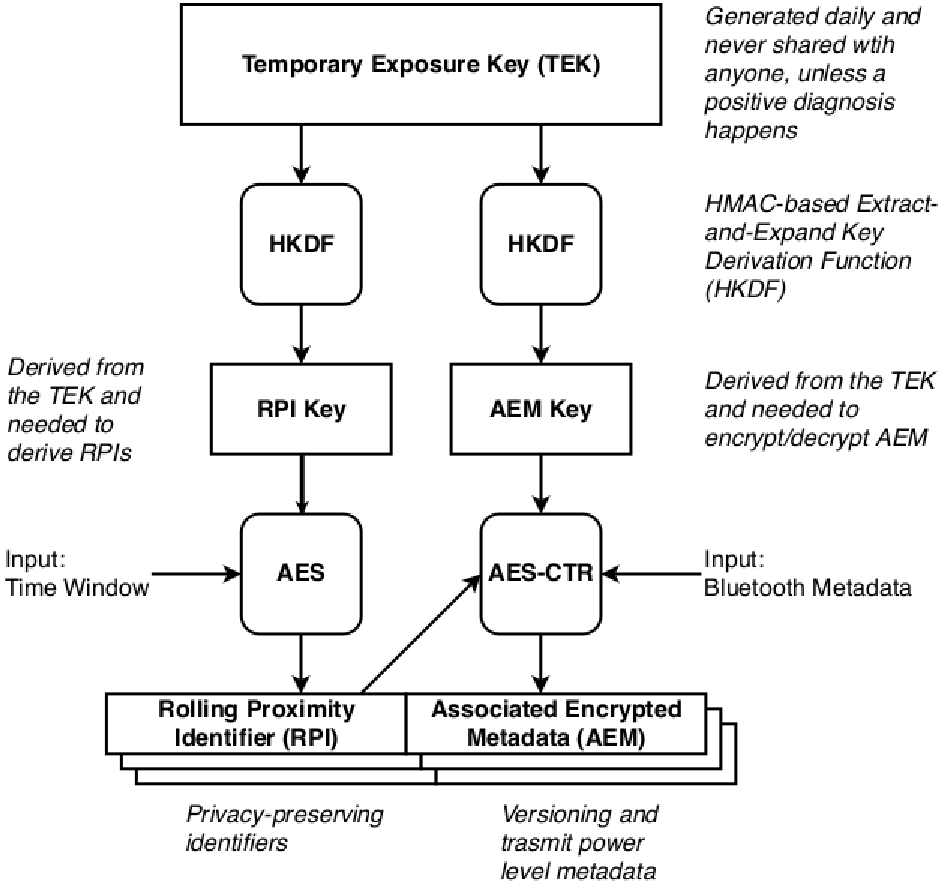}
  \caption{GAEN - Key Generation Workflow.}
  \label{fig:gaencrypto}
\end{figure}

\textbf{Pseudonym Generation and Exchange.} Once a \gls{GAEN}-based app is installed on a device, and the onboarding process is completed, it generates a 16-byte \gls{TEK}, which is valid for a single day and then replaced by a new one. As shown in Figure~\ref{fig:gaencrypto}, a \gls{GAEN}-based app generates a 16-byte \gls{RPIK} and an \gls{AEMK} from its current \gls{TEK}. Finally, the app derives the \gls{RPI} from the \gls{RPIK}. Since RPI are not linked to a specific person or device, they can be exchanged as pseudonyms during contact tracing. Every time the smartphone \gls{BLE} MAC randomized address changes, the current \gls{RPI} needs to be updated accordingly, and a new one is derived again from the \gls{RPIK}, with a two-hour validity windows. Advertisement packets are broadcasted by the app via \gls{BLE}, and stored locally by other nearby apps.

\textbf{Announcement of a New Positive Person.} When a new user is found positive to SARS-CoV-2, he can choose to upload his recent \gls{TEK} (also referred as Diagnosis Key) to the centralized server. Other apps periodically download the \gls{TEK} of newly infected users, infer the associated \gls{RPI} and check if those \gls{RPI} match with the ones from the app's own recent contacts. If a match is found, the app uses the \gls{AEMK} to decrypt the \gls{AEM} and to evaluate the transmission signal strength, which contributes to the final risk score, alongside with the duration of the contact.

\subsection{Threat Model}
\label{ssec:threat_model}
\begin{figure}[h!]
  \centering
  \includegraphics[width=\linewidth]{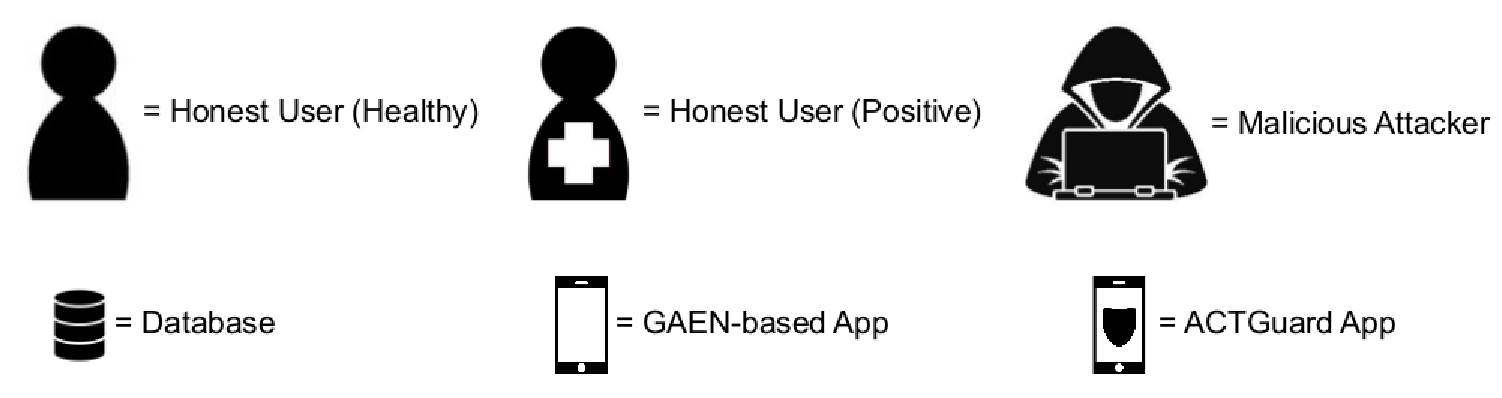}
  \caption{Symbols Explanation.}
  \label{fig:legend}
\end{figure}

In our threat model, as well as in the scenarios described in Section~\ref{sec:ACTGuard_design}, we assume to have the components shown in Fig.~\ref{fig:legend}: a set of three honest users, among which one is positive to SARS-CoV-2; two malicious attackers; a set of servers, which could be either the health authority one or the \emph{ACTGuard} one; a \gls{GAEN}-based app; the \emph{ACTGuard} app.

\begin{figure}[h!]
  \centering
  \includegraphics[width=\linewidth]{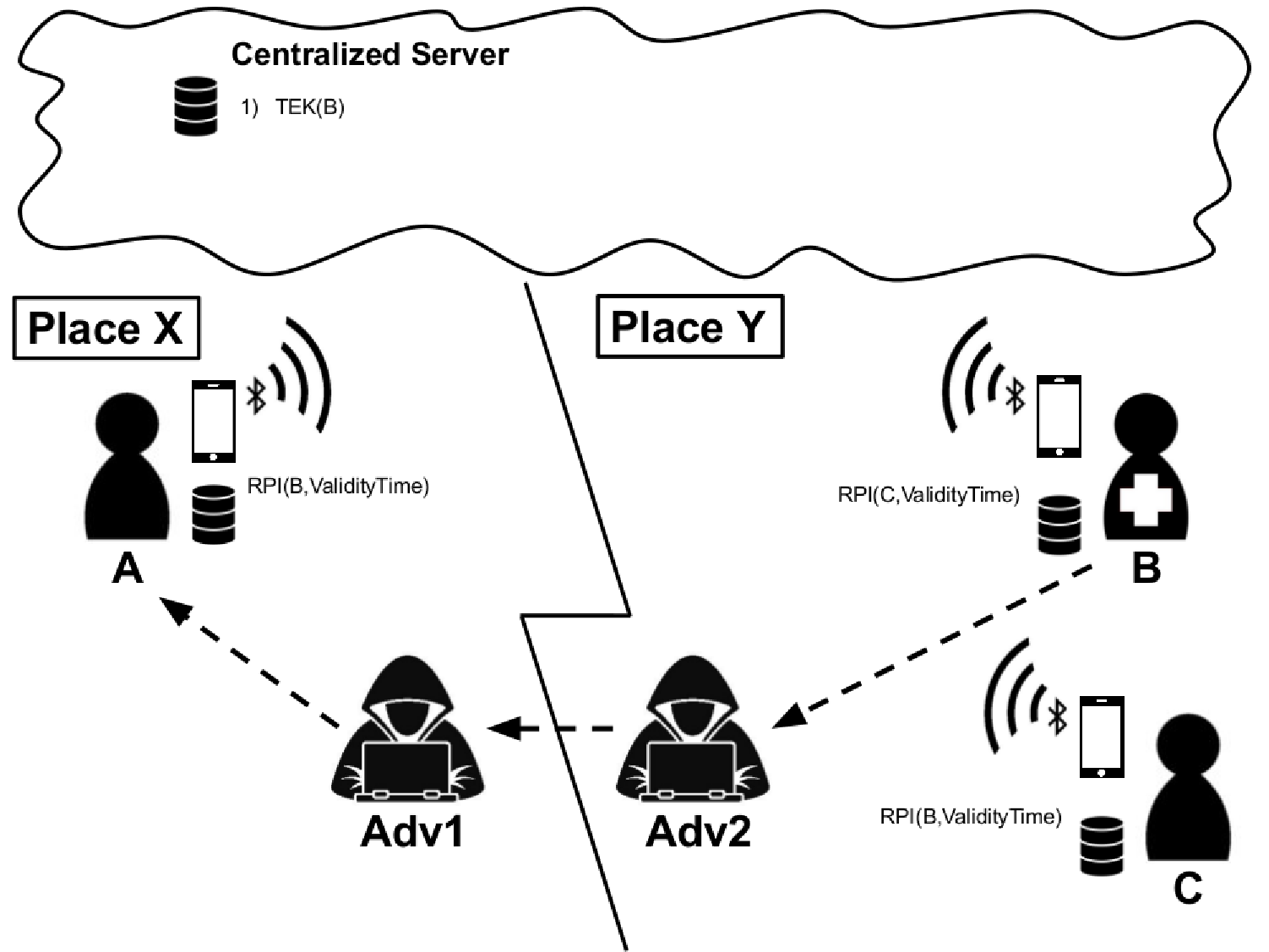}
  \caption{GAEN Relay Attack.}
  \label{fig:s0}
\end{figure}

Our relay attack against \gls{GAEN}-based apps utilizes the threat model shown in Fig.~\ref{fig:s0}. It involves the actors detailed in Table~\ref{tab:actors0}: three honest users (i.e., A, B and C), with an active \gls{GAEN}-based app on their smartphone; two adversaries (i.e., Adv1 and Adv2) without any \gls{GAEN}-based app on their smartphones. 

\begin{table}
	\centering
    \begin{tabular}{lcc}
        \toprule
        Actor & Alignment & GAEN-based App \\
        \midrule
        Alice & Honest & \cmark \ \\
        Bob & Honest & \cmark \\
        Carlos & Honest & \cmark \\
        Adv1 & Malicious & \xmark \\
        Adv2 & Malicious & \xmark \\
        \bottomrule
    \end{tabular}
    \vspace*{1mm}
    \caption{GAEN Relay Actors.}
    \label{tab:actors0}
\end{table}

In this scenario, we assume that A and Adv1 are positioned in Place X, while B, C and Adv2 are positioned in Place Y. In our threat model, we assume that Adv2 stands near B and relays his advertising data to Adv1. In particular, Adv2 aims at stealing the signal transmitted by B, who is going to be found positive in a near future. Once acquired, Adv2 shares such signal with Adv1, which starts advertising it in a different location, exposing the victims to an infected pseudonym, eventually inflating their risk score and creating a false positive case. The attack is performed as described below. 

\textbf{Proximity Tracing.} A meets Adv1 in Place X: GAEN-App(A) broadcasts \gls{BLE} advertising packets containing his current RPI, i.e., RPI(A,ValidityTime). By exploiting the signal intercepted by Adv2, Adv1 maliciously broadcasts advertising packets containing RPI(B,ValidityTime), which is locally stored by the GAEN-App(A). Similarly, B meets C in Place Y: GAEN-App(B) advertises RPI(B,ValidityTime), while storing RPI(C,ValidityTime), and GAEN-App(C) does it the other way around. 

\textbf{Positive Diagnosis.} At some point, B gets diagnosed as positive to SARS-CoV-2 and uploads his past \gls{TEK} to the centralized server, as they are needed to retrieve the corresponding \gls{RPI}. GAEN-App(C) finds a match with RPI(B,ValidityTime). Consequently, C is notified about the new exposure. In this case, GAEN-App(C) is working as intended. However, GAEN-App(A) finds a match with RPI(B,ValidityTime), too. Consequently, A is notified about the new exposure. However, in reality, A only met Adv1, that was impersonating B during the relay attack. In this case, GAEN-App(A) is not working as intended.

\textbf{Relay Attack Impact.} Relay attacks make \gls{ACT} apps significantly less reliable, as they can inflate health risk scores by creating contacts that have never occurred. Adversaries performing relay attacks can choose sensible locations: Place Y could be a location with a high exposure risk (i.e. a hospital, a quarantined city), whereas Place X could be a location with many individuals coming and going (i.e. a train station, an airport). Apart from inflating health risk scores, the attackers might also push false positive warnings to specific individuals, instead of targeting whole groups. Possible attack scenarios could be the following ones: a student enforcing a professor to the self-quarantine, thus leading to the cancellation of an important exam; the trading of pseudonyms derived from infected people in the dark net; the real-time monitoring of a person movements by locating several devices in places often visited by the targeted victim. 
\section{\emph{ACTGuard} Design}
\label{sec:ACTGuard_design}

\gls{ACT} apps using \gls{BLE} are vulnerable to relay attacks by design: proximity tracing protocols depend on the close distance between two app users, but malicious attackers can intercept and relay \gls{BLE} signals. Solutions that implement authentication mechanisms, or rely on location or timing data have limitations: integrating authentication mechanisms in the communication protocol introduces privacy issues; location data can be used to validate the distance between users, but it can not be shared with anyone else, since it would still cause privacy issues; timing data could be useful, but \gls{RPI} have a two-hour validity window, which could be long enough for an attacker to execute a relay attack. 

We designed \emph{ACTGuard}, considering the following two objectives: 
\begin{enumerate}
    \item enabling \gls{GAEN}-based apps to defend against relay attacks. 
    \item guaranteeing the same privacy level as the current \gls{GAEN}-based apps.
\end{enumerate}

Our intuition behind the design of \emph{ACTGuard} relies on the collection of location data related to a contact between two app users. More specifically, by saving the GPS coordinates of a contact between two users, we can detect when \gls{RPI} are transmitted at the same time in two different places. This method prevents the delivery of false positive alerts to victims that received relayed \gls{RPI}. The design of \emph{ACTGuard} requires the app user to save the following data for each contact occurred between user A and user B:
\\\\
{\small
hash\{ \texttt{RPI(A,ValidityTime)}, \texttt{RPI(B,ValidityTime)}, \texttt{Location}, \texttt{ContactTime} \}}

To avoid any inconsistency, RPI are ordered alphabetically. \emph{ACTGuard} saves the set of hashes associated to contacts involving its owner. If any user is found infected, he uploads his past hashes to a central server. Other \emph{ACTGuard} users periodically download the hashes of the infected people and check them against the ones they saved locally. If all the information concerning a contact between two users (i.e., both RPI, location, time) is equal, then hash values will perfectly match and \emph{ACTGuard} will confirm that the contact occurred for real.
If this does not happen, there are two possible reasons:
\begin{enumerate}
    \item the infected user app did not upload the hashes to the centralized server;
    \item the infected user uploaded the hashes to the centralized server, and a relay attack was performed.
\end{enumerate}
Besides preventing relay attacks, \emph{ACTGuard} provides also a privacy-preserving solution, since each user only shares hashes. Thus, the centralized server is able to infer none of the contact information as hash functions are not reversible by design. 

To better explain the \emph{ACTGuard} resilience to relay attacks, we consider our threat model, which is illustrated in Fig.~\ref{fig:s0}, and analyze several scenarios that differ according to which actors are equipped with \emph{ACTGuard}. In particular, we assume that:
\begin{itemize}
    \item A, B and C always use a \gls{GAEN}-based app.
    \item Adv1 and Adv2 use neither a \gls{GAEN}-based app nor \emph{ACTGuard}.
    \item A always uses \emph{ACTGuard}, since he is the victim and he has to defend against relay attacks. 
    \item B and C might not use \emph{ACTGuard}.
\end{itemize}

With the above-mentioned assumptions, we identified the following four scenarios:
\begin{itemize}
    \item Scenario 1 - A, B and C all using \emph{ACTGuard}. 
    \item Scenario 2 - A and C use \emph{ACTGuard}, but B does not. 
    \item Scenario 3 - A and B use \emph{ACTGuard}, but C does not. 
    \item Scenario 4 - A uses \emph{ACTGuard}, but B and C do not. 
\end{itemize}

In the rest of the section, we will discuss Scenario 1 (i.e, Section~\ref{ssec:scenario1}) and Scenario 2 (i.e, Section~\ref{ssec:scenario2}). Since C is not infected, thus not going to uploaded his contact information, his adoption of \emph{ACTGuard} does not affect the defence against the relay attack. Thus, we do not discuss Scenario 3. Similarly, the focus of the Scenario 4 would be B not having \emph{ACTGuard}, which is already analyzed in Scenario 2.

\subsection{Scenario 1: All Users Having \emph{ACTGuard}}
\label{ssec:scenario1}
\begin{figure}[h!]
  \centering
  \includegraphics[width=\linewidth]{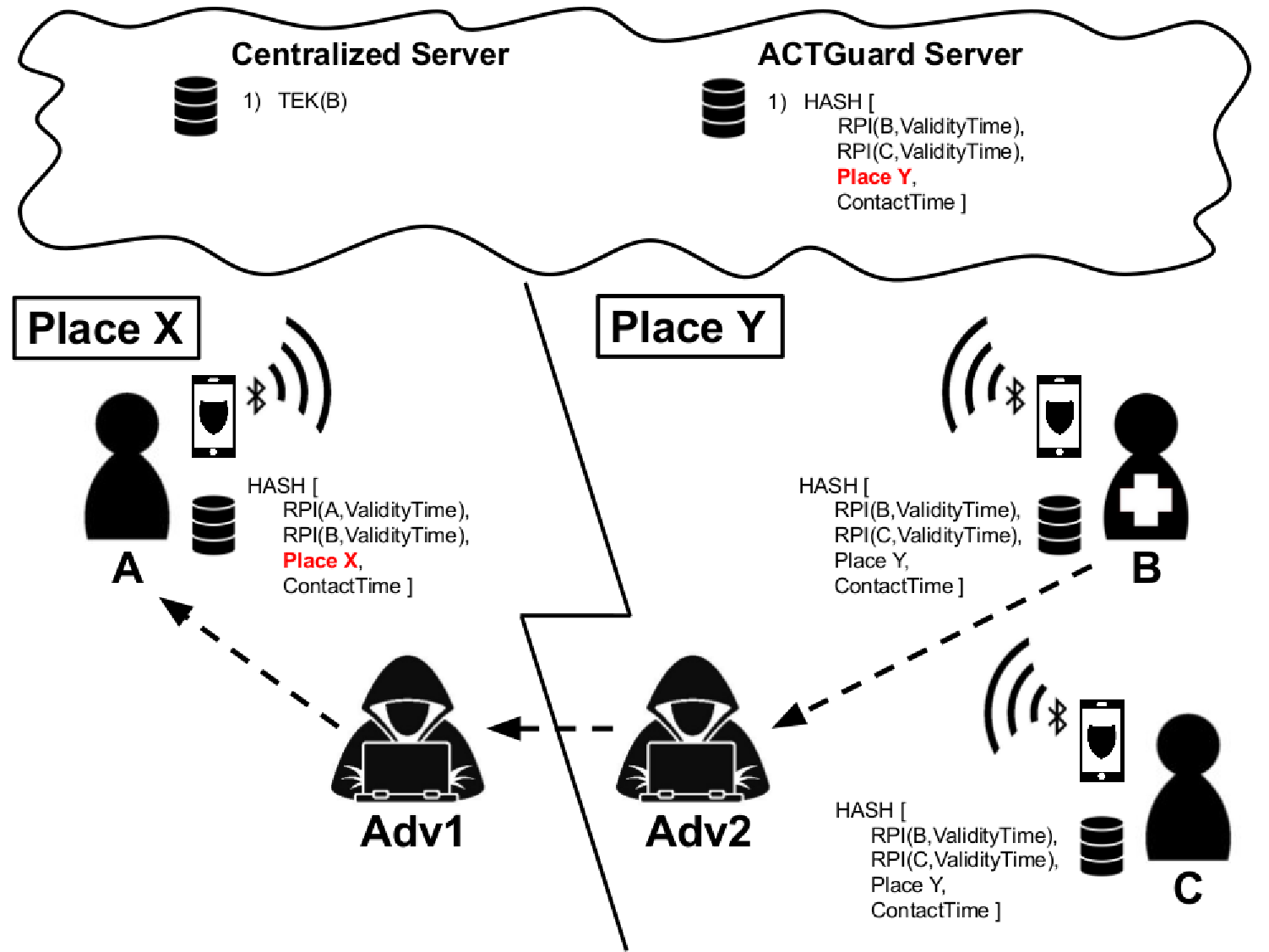}
  \caption{Scenario 1 - All Users Having \emph{ACTGuard}.}
  \label{fig:s1}
\end{figure}

In Scenario 1, we assume that all honest users (i.e., A, B and C) are equipped with a \gls{GAEN}-based app and \emph{ACTGuard}, while the attackers (i.e., Adv1 and Adv2) have none of them. Scenario 1 is illustrated in Fig.~\ref{fig:s1} and the details concerning the different actors are summarized in Table~\ref{tab:actors1}.

\begin{table}[ht!]
    \centering
    \begin{tabular}{lccc}
        \toprule
        Actor & Alignment & GAEN-based App & ACTGuard \\
        \midrule
        A & Honest & \cmark & \cmark \\
        B & Honest & \cmark & \cmark \\
        C & Honest & \cmark & \cmark \\
        Adv 1 & Malicious & \xmark & \xmark \\
        Adv2 & Malicious & \xmark & \xmark \\
        \bottomrule
    \end{tabular}
    \vspace*{1mm}
    \caption{Actors of Scenario 1 - All Users Having \emph{ACTGuard}.}
    \label{tab:actors1}
\end{table}

\textbf{Contact.} A meets Adv1 in Place X: GAEN-App(A) is advertising RPI(A,ValidityTime) and Adv1 is maliciously advertising RPI(B,ValidityTime). GAEN-App(A) stores RPI(B,ValidityTime), while ACTGuard(A) stores the hash of the contact (i.e., RPI(A,ValidityTime), RPI(B,ValidityTime), Location, ContactTime). Similarly, B meets C in Place Y: GAEN-App(B) is advertising RPI(B,ValidityTime) and GAEN-App(C) is advertising RPI(C,ValidityTime). Consequently, GAEN-App(B) stores RPI(C,ValidityTime) and GAEN-App(C) stores RPI(B,ValidityTime). At the same time, ACTGuard(B) and ACTGuard(C) both store the exact same hash (i.e., RPI(B,ValidityTime), RPI(C,ValidityTime), Location, ContactTime). 

\textbf{Positive Diagnosis.} We assume that B gets diagnosed as positive to SARS-CoV-2 and through his \gls{GAEN}-based app, he uploads his past \gls{TEK} to the centralized server. B and C had a real contact. Thus, GAEN-App(C) downloads TEK(B), derives all possible \gls{RPI} and finds a match with RPI(B,ValidityTime) saved in its local storage. Consequently, C is notified about a potential health risk. In this case, GAEN-App(C) is working as intended. At the same time, GAEN-App(A) finds a match with the RPI(B,ValidityTime) saved in the local storage, thus leading to a potential health risk warning, as well. However, in reality, A only met Adv1, while he was impersonating B during the relay attack. In this case, GAEN-App(A) is not able to discriminate between a real contact and a forged one. Meanwhile, ACTGuard(C) finds a correct match and it confirms the health warning from GAEN-App(C). On the contrary, ACTGuard(A) does not find a match in the hashes due to the different location of the contacts. Consequently, ACTGuard(A) does not confirm the health risk warning from GAEN-App(A). Thanks to \emph{ACTGuard}, the relay attack against A is detected and a false positive health risk warning is prevented. 

\subsection{Scenario 2: Positive Person without \emph{ACTGuard}}
\label{ssec:scenario2}
\begin{figure}[h!]
  \centering
  \includegraphics[width=\linewidth]{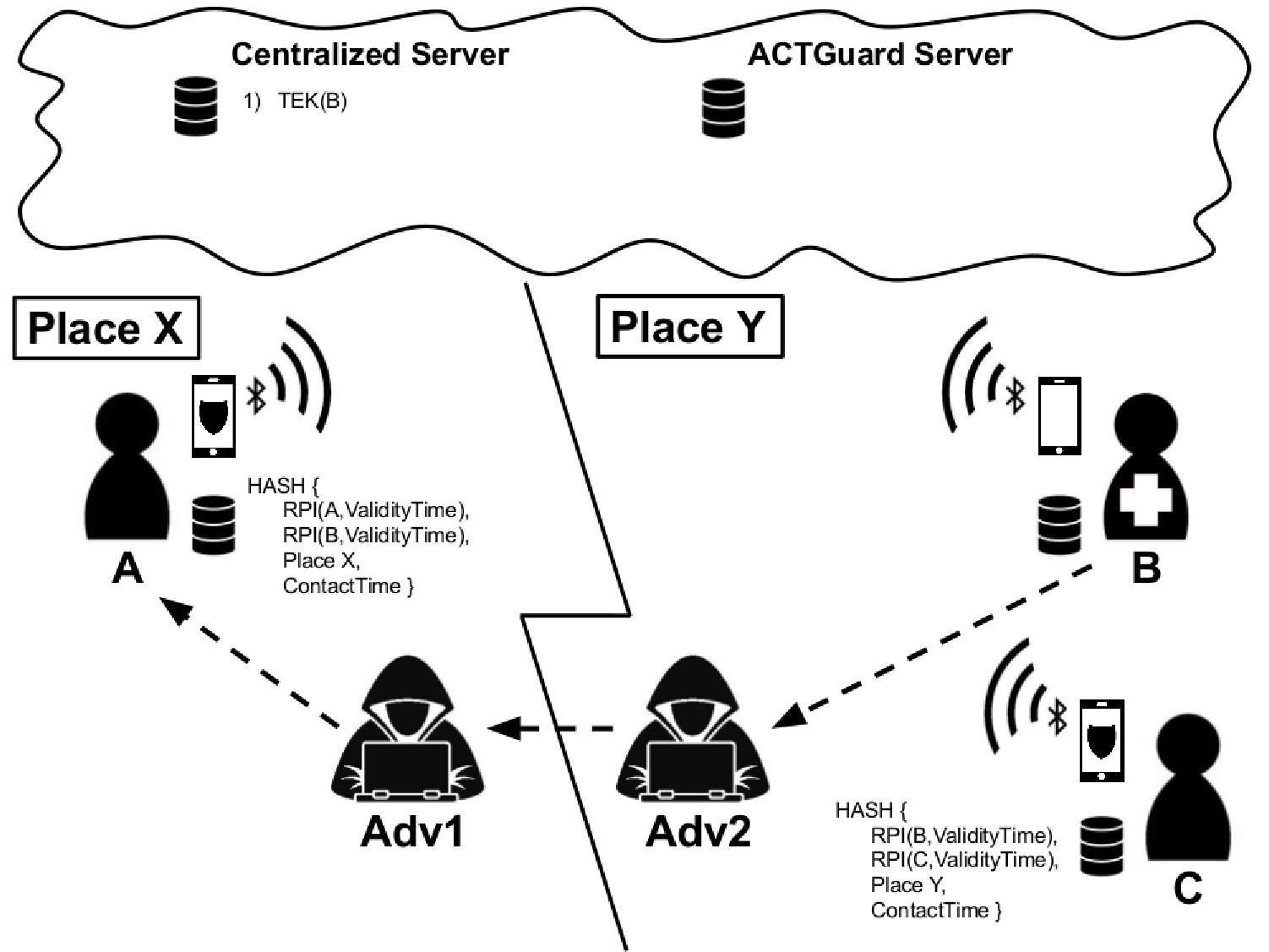}
  \caption{Scenario 2 - Positive Person without \emph{ACTGuard}.}
  \label{fig:s2}
\end{figure}

In Scenario 2, we assume that the honest users A and C are equipped with a \gls{GAEN}-based app and \emph{ACTGuard}, that B is only a \gls{GAEN}-based app user and that the attackers (i.e., Adv1 and Adv2) have none of them. Scenario 2 is illustrated in Fig.~\ref{fig:s2} and the details concerning the different actors are summarized in Table~\ref{tab:actors2}.

\begin{table}[ht!]
	\centering
    \begin{tabular}{lccc}
        \toprule
        Actor & Alignment & GAEN-based App & ACTGuard \\
        \midrule
        A & Honest & \cmark & \cmark \\
        B & Honest & \cmark & \xmark \\
        C & Honest & \cmark & \cmark \\
        Adv 1 & Malicious & \xmark & \xmark \\
        Adv2 & Malicious & \xmark & \xmark \\
        \bottomrule
    \end{tabular}
    \vspace*{1mm}
    \caption{Actors of Scenario 2 - Positive Person without \emph{ACTGuard}.}
    \label{tab:actors2}
\end{table}

\textbf{Contact.} A meets Adv1 in Place X: GAEN-App(A) is advertising RPI(A,ValidityTime) and Adv1 is maliciously advertising RPI(B,ValidityTime). GAEN-App(A) stores RPI(B,ValidityTime), while ACTGuard(A) stores the hash of the contact (i.e., RPI(A,ValidityTime), RPI(B,ValidityTime), Location, ContactTime). Similarly, B meets C in Place Y: GAEN-App(B) is advertising RPI(B,ValidityTime) and GAEN-App(C) is advertising RPI(C,ValidityTime). Consequently, GAEN-App(B) stores RPI(C,ValidityTime) and GAEN-App(C) stores RPI(B,ValidityTime). At the same time, ACTGuard(C) stores the hash of the contact (i.e., RPI(B,ValidityTime), RPI(C,ValidityTime), Location, ContactTime).

\textbf{Positive Diagnosis.} We assume that B gets diagnosed as positive to SARS-CoV-2 and through his \gls{GAEN}-based app he uploads his past \gls{TEK} to the centralized server. B and C had a real contact. Thus, GAEN-App(C) downloads TEK(B), derives all possible \gls{RPI} and finds a match with RPI(B,ValidityTime) saved in its local storage. Consequently, C is notified about a potential health risk. In this case, GAEN-App(C) is working as intended. At the same time, GAEN-App(A) finds a match with the RPI(B,ValidityTime) saved in the local storage, thus leading to a potential health risk warning, as well. However, in reality, A only met Adv1, while he was impersonating B during the relay attack. In this case, GAEN-App(A) is not able to discriminate between a real contact and a forged one. Both ACTGuard(A) and ACTGuard(C) do not find a match in the hashes shared by B, since B does not use \emph{ACTGuard} and did not upload any hash. Since ACTGuard(A) is not able to confirm the health risk warning from GAEN-App(A), it can not confirm if a relay attack actually happened.

The Scenario 2 opens up two further situations: 
\begin{itemize}
    \item A use case: A met Adv1 while he was performing the relay attack. Since B does not use \emph{ACTGuard}, A can not have any confirmation about the health risk alert received after the contact with the impersonator of B.
    \item C use case: C really met B. However, since B does not use \emph{ACTGuard}, C can not have any confirmation about the health risk alert received after the contact with the real B.
\end{itemize}
\section{Use Case Study: the Italian Contact Tracing App Immuni}
\label{sec:immuniguard_design}
In this section, we describe the \gls{ACT} app we chose as a case study to develop a proof of concept of the relay attack and of the \emph{ACTGuard} design. In particular, in Section~\ref{ssec:immuni_overview}, we provide an overview of \emph{Immuni}~\cite{ct-immuni}, the Italian official \gls{ACT} app. In Section~\ref{ssec:relay_attack_immuni}, we describe the relay attack we developed against \emph{Immuni} and, finally, in Section~\ref{ssec:ACTGuard_implementation}, we illustrate the implementation details of \emph{ImmuniGuard}, which is an instance of \emph{ACTGuard} referring to the \emph{Immuni} use case. 

\subsection{\emph{Immuni} Overview}
\label{ssec:immuni_overview}

\emph{Immuni} a mobile contact tracing app based on the \gls{GAEN} protocol~\cite{web-gaen}, thus adhering to the \emph{decentralized} approach and relying on the \gls{BLE} technology for proximity tracing.

\textbf{Onboarding.} Once installed, \emph{Immuni} detects the user language, checks for required updates and provides a basic explanation of how \emph{Immuni} works. Then, \emph{Immuni} shows the privacy note and the user agreement, asking if the user is more than fourteen years old. At last, the user submits his province of residence and enables the required app permissions.

\emph{Immuni} interacts with a back-end architecture involving the following services:
\begin{itemize}
    \item Exposure Ingestion Service.
    \item Exposure Reporting Service.
    \item Backend \gls{OTP} Service.
    \item Analytics Service.
    \item App Configuration Service.
\end{itemize}

\textbf{Exposure Ingestion Service.} The Exposure Ingestion Service provides a set of \gls{API} for \emph{Immuni} apps to enable the upload of \gls{TEK} generated over the past 14 days, when an app user is found infected and he is willing to share his keys. Contextually, the infected user uploads the epidemiological information from the previous 14 days. If any epidemiological information is indeed uploaded, the user province of domicile is uploaded, too. This process can only take place with an authorised \gls{OTP} from the Backend \gls{OTP} Service. The Exposure Ingestion Service is also responsible for the periodical generation of \gls{TEK} chunks, to be published by the Exposure Reporting Service. The \gls{TEK} chunks are assigned a unique incremental index and are immutable. The province of domicile and the epidemiological information are forwarded to the Analytics Service.

\textbf{Exposure Reporting Service.} The Exposure Reporting Service makes the \gls{TEK} chunks created by the Exposure Ingestion Service available to other apps. Only TEK chunks from the past 14 days are made available.

\textbf{Backend \gls{OTP} Service.} The Backend \gls{OTP} Service provides a set of \gls{API} to the National Healthcare Service for authorising \gls{OTP} that can be used to upload data from \emph{Immuni} apps via the Exposure Ingestion Service. \emph{Immuni} generates the \gls{OTP}, that the app user personally communicates to a healthcare operator. Then, the healthcare operator inserts the \gls{OTP} into the Italian Health Information System, registering it on the Backend \gls{OTP} Service. The \gls{OTP} automatically expires after a defined time period.

\textbf{Analytics Service.} The Analytics Service provides a set of \gls{API} for \emph{Immuni} apps to upload data without identifying users, both during regular operations and especially when a match is found between \gls{TEK} chunks and \gls{RPI}. To ensure the proper functioning of the system, maximize the effectiveness of the exposure notifications and provide an optimal healthcare assistance to users, the following information is sent to the server:
\begin{itemize}
    \item TEK.
    \item Epidemiological information.
    \item Operational information.
\end{itemize}

TEK are required to allow other \emph{Immuni} users to calculate their risk of being positive to SARS-CoV-2. Whenever TEK are uploaded, the following epidemiological information is uploaded, as well:
\begin{itemize}
    \item The day the exposure occurred.
    \item The duration of the exposure.
    \item The signal attenuation information used for estimating the distance between the two users’ devices during the exposure.
\end{itemize}

Whenever an exposure detection has been completed, the following operational information is uploaded:
\begin{itemize}
    \item Whether the device runs iOS or Android.
    \item Whether permission to leverage the \gls{GAEN} protocol is granted.
    \item Whether the device \gls{BLE} is enabled.
    \item Whether permission to send local notifications is granted.
    \item Whether the user was notified of a risky exposure after the last exposure detection (i.e., after the app has downloaded new \gls{TEK} from the server and detected if the user has been exposed to users positive to SARS-CoV-2).
    \item The date on which the last risky exposure took place, if any.
\end{itemize}

Along with genuine analytics uploads, after every exposure detection event, \emph{Immuni} apps may perform dummy analytics uploads, indistinguishable from the genuine ones.

\textbf{App Configuration Service.} The App Configuration Service updates the Configuration Settings every time the app starts a new background or foreground session. Such settings can be used for tuning traffic-analysis mitigation measures and update weights used in the risk prediction model.

\subsection{Relay Attack against \emph{Immuni}}
\label{ssec:relay_attack_immuni}

Here, we illustrate the workflow of the relay attack we designed against \emph{Immuni}. Our scenario involves the following actors: 
\begin{itemize}
    \item Immuni App 1: this is the first victim of the relay attack, since its \gls{BLE} packet is sniffed and re-transmitted by the attacker. 
    \item Malicious App 1: this is the first malicious actor, responsible for sniffing the \gls{BLE} packet, which will be re-transmitted afterwards.
    \item Malicious Database: this database is used by the two attackers to share the sniffed  \gls{BLE} packet, even though they are located in different places. 
    \item Malicious App 2: this is the second malicious actor, responsible for the re-transmission of the sniffed \gls{BLE} packet. 
    \item Immuni App 2: this is the second victim of the relay attack, since it receives the sniffed \gls{BLE} packet.  
\end{itemize}

\begin{figure*}[h!]
  \centering
  \includegraphics[width=\linewidth]{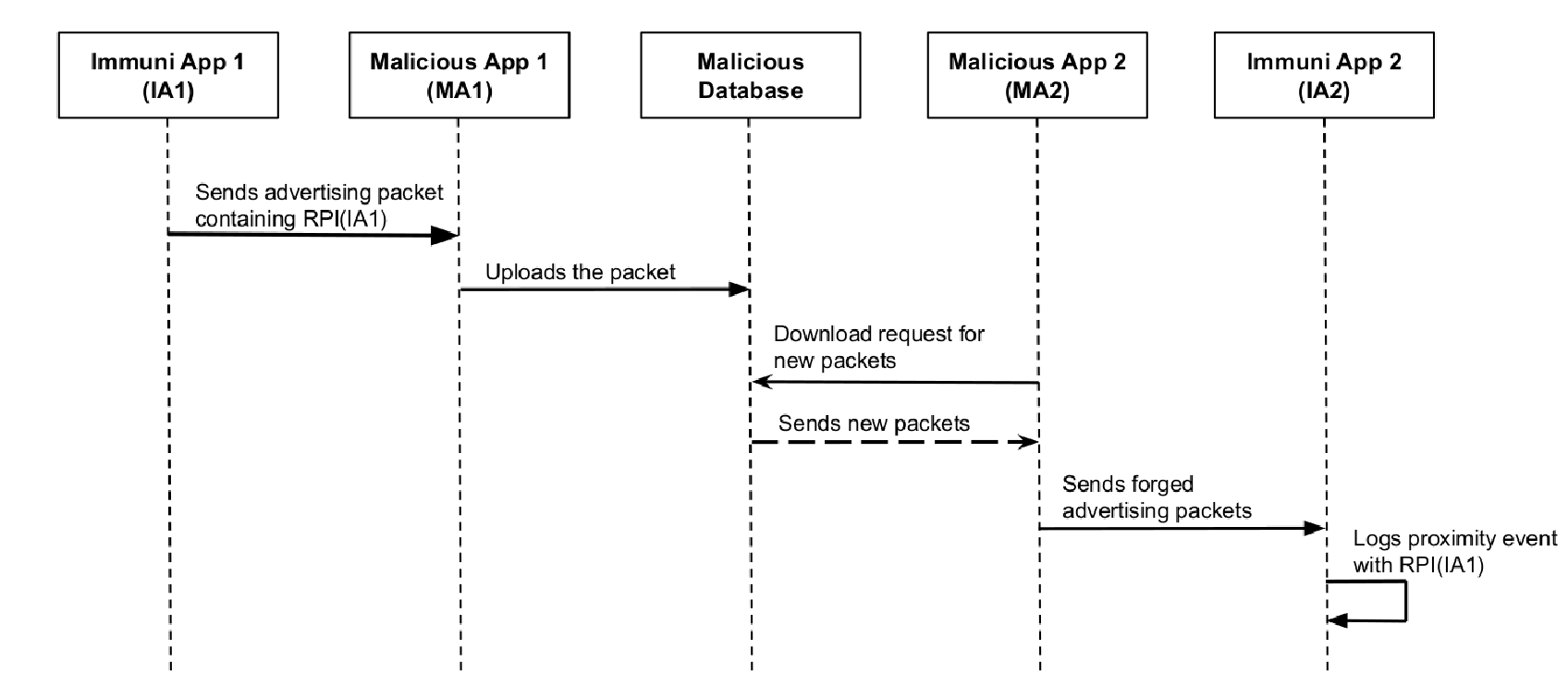}
  \caption{Workflow of the Relay Attack we Designed against \emph{Immuni}.}
  \label{fig:relayimmuni}
\end{figure*}

As shown in Fig.~\ref{fig:relayimmuni}, Immuni App 1 keeps advertising its \gls{BLE} packets, which contain its current \gls{RPI}. Malicious App 1 is located nearby the Immuni App 1, thus being able to intercept its advertising packets and send them to the Malicious Database. Meanwhile, Malicious App 2 downloads the sniffed \gls{BLE} packets from the Malicious Database and starts advertising them, pretending to be Immuni App 1. Finally, Immuni App 2 becomes the victim of the relay attack by receiving the sniffed \gls{BLE} packets from Malicious App 2. 

Even though our implementation refers to the \emph{Immuni} use case, the above-mentioned relay attack can be applied on any \gls{GAEN}-based app. Since such apps do not track the user location, they are not able to detect relay attacks. Attackers can broadcast sniffed packets from any location, pretending to be the original owner of the \gls{BLE} packet, as soon as it is re-transmitted within the time window of about two hours.

\subsection{\emph{ImmuniGuard}: An \emph{ACTGuard} Implementation for \emph{Immuni}}
\label{ssec:ACTGuard_implementation}
Our implementation of the \emph{ACTGuard} design is called \emph{ImmuniGuard} and it was developed considering the Italian \gls{ACT} app, i.e., \emph{Immuni}. We implemented \emph{ImmuniGuard} as an Android app, requiring an Android SDK version equal to 29. The experiments were performed on a Xiaomi Redmi 5 Plus and a complete demo video of the defence applied by \emph{ImmuniGuard} against a relay attack is available online\footnote{\url{https://spritz.math.unipd.it/projects/immuniguard/demo/immuniguard-demo.avi}}. 

\emph{ImmuniGuard} requires the following permissions:
\begin{itemize}
    \item \texttt{INTERNET} and \texttt{INTERNET\_NETWORK\_STATE}, to interact with an online database.
    \item \texttt{BLUETOOTH} and \texttt{BLUETOOTH\_ADMIN}, for Bluetooth scanning and advertising features.
    \item \texttt{ACCESS\_FINE\_LOCATION}, for GPS tracking (can be downgraded to \texttt{ACCESS\_COARSE\_LOCATION}).
    \item \texttt{FOREGROUND\_SERVICE}, for regular monitoring of GPS locations.
\end{itemize}
Below, we will go through the main steps involved in the \emph{ImmuniGuard} workflow. 

\textbf{Scanning for \emph{Immuni} advertisement packets.} \emph{ImmuniGuard} uses the standard \gls{BLE} library, provided by the Android platform, to regularly perform scans. During the scans, \emph{ImmuniGuard} specifically looks for \emph{Immuni} advertisement packets, identified by a value equal to "0xFD6F" in the Service UUID, as any other \gls{GAEN}-based app.

\textbf{Storing RPIs.} For each contact with another \emph{Immuni} app user, \emph{ImmuniGuard} saves the advertised RPI, the current location and the current time. This data is stored in two databases:
\begin{itemize}
    \item MyContactsTable, containing information related to contacts with other \emph{Immuni} app users.
    \item PositiveTable, storing information uploaded by positive users to \emph{ImmuniGuard} database.
\end{itemize}
New entries are added to MyContactsTable every time a new contact occurs, including their hash value. PositiveTable is updated whenever \emph{ImmuniGuard} tries to download new information from the \emph{ImmuniGuard} online database. \emph{ImmuniGuard} should also store the RPIs broadcasted by the app user \emph{Immuni} app, but this information is currently restricted. In our current proof of concept, \emph{ImmuniGuard} randomly generates its own dummy RPI, unrelated to \emph{Immuni}.

\textbf{Uploading data if infected.} Whenever \emph{ImmuniGuard} users are diagnosed as SARS-CoV-2 positive, they can upload the content of the MyContactsTable database, locally stored on the app. In a real scenario, the data upload should be validated by public health authorities. In our proof of concept, this functionality can be performed anytime. 

\textbf{Periodically downloading new infected data.} \emph{ImmuniGuard} needs to regularly download information about new users diagnosed as SARS-CoV-2 positive. In a real scenario, this process would be automated, but in our proof of concept, this functionality can be performed anytime. 

\textbf{Updating health risk status.} The health risk status is calculated by comparing and matching the hashes in MyContactsTable and in PositiveTable, to find if the owner of \emph{ImmuniGuard} was in proximity of an infected user. In our proof of concept, the health risk status is represented by the number of infected contacts.

\subsection{\emph{ImmuniGuard} Limitations}

\emph{ImmuniGuard} is a standalone app, and it is not allowed to access to the list of RPI broadcasted by the active instance of \emph{Immuni} app installed on the user's device. \emph{ImmuniGuard} needs that list, in order to compute its own contact hashes, store them inside MyContactsTable and compare them with the infected hashes downloaded from the \emph{ImmuniGuard} server. Otherwise, the detection of relay attacks is not possible. This issue can be easily solved with the integration of our solution as a feature of the official app, thus gaining direct access to the list of RPI broadcasted by \emph{Immuni}.
\section{Related Work}
\label{sec:related_work}

In this section, we describe in details the \gls{ACT} apps we analyzed (i.e., Section~\ref{ssec:act_apps}), specifically focusing on the privacy (i.e., Section~\ref{ssec:privacy_issues_act_apps}) and security (i.e., Section~\ref{ssec:sec_issues_act_apps}) issues that affect them. 

\subsection{Contact Tracing Apps}
\label{ssec:act_apps}

\textbf{TraceTogether.} TraceTogether~\cite{web-bluetrace} is the first contact tracing app based on \gls{BLE} technology and developed in Singapore, one of the first countries affected by SARS-CoV-2. It implements the BlueTrace protocol, exchanging temporary identifiers through \gls{BLE} advertisement packets and allowing user to willingly upload them on a centralized health authority server. During interviews with patients positive to SARS-CoV-2, medical operators ask if the patient installed TraceTogether and are able to dynamically adjust proximity and duration filtering thresholds for a better contact tracing performance.

\textbf{Safe Paths.} Safe Paths~\cite{rel-rogue} is a secure location logging technology, based on the "Private Automated Contact Tracing (PACT)" protocol, developed by MIT researchers. It currently supports only GPS tracking, with plans for other location and proximity technologies. Safe Paths stores the GPS movement trails of its owner, who can willingly share them with an authorized public health authority as part of the contact tracing process.

\textbf{StayHomeSafe.} StayHomeSafe~\cite{ct-stayhomesafe} is a mobile app developed by the Government of the Hong Kong Special Administrative Region. The Hong Kong procedure for compulsory quarantine requires wearing a wristband connected to the StayHomeSafe app. The app utilizes an innovative geofencing technology. The initial setup requires the user to slowly walk around the house so that the app can record data signals (e.g., Wi-Fi, Bluetooth, GPS) as a form of unique signature of that house, and detects whenever the user leaves this virtual fence. StayHomeSafe does not collect sensitive data, nor discloses the location of its user.

\textbf{Hamagen.} Hamagen~\cite{ct-hamagen} is a mobile app developed by the Israel Ministry of Health, which employs both GPS and \gls{BLE} technology. Hamagen regularly saves the device location coordinates and compares them with the GPS trails of infected people. Additionally, it can perform proximity tracing through \gls{BLE}. All data, which the Ministry of Health has access to, is explicitly shared by infected patients.

\textbf{Aarongya Setu.} Aarongya Setu~\cite{ct-aarongyasetu} is a mobile app developed by the Government of India. The app is mandatory for all Indian citizens, and it is also used by employers to monitor the health risk status of their employees. Aarongya Setu uses both GPS tracking and \gls{BLE} technology. During the initial setup, the user is required to submit personal data such as his/her name and phone number.

\textbf{StopCovid.} StopCovid~\cite{ct-stopcovid} is a mobile app developed by the National Institute for Research in Digital Science and Technology (INRIA), as requested by the Government of France. The app uses \gls{BLE} technology, and implements the ROBERT protocol, as the Government considers a centralized approach more secure.

\textbf{SwissCovid.} SwissCovid~\cite{ct-swisscovid} is a mobile app developed by the Switzerland Federal Office of Public Health. The app uses \gls{BLE} technology, GAEN in particular, to exchange anonymous temporary identifiers. Whenever a user is infected, he can willingly reveal and submit his temporary identifiers in order to warn other app users met in the past days.

\textbf{Covid Alert NY.} Covid Alert NY~\cite{ct-alertny} is a mobile app commissioned by the New York State Department of Health. Akin to many other American contact tracing apps, it implements GAEN. A special feature allows users to willingly disclose personal information, anonymously, such as the user county, gender, age-range, ethnicity, and symptoms. Technical data can be also be anonymously shared with the centralized server, if the user chooses to do so.

\subsection{Privacy Issues of Contact Tracing Apps}
\label{ssec:privacy_issues_act_apps}

The primary purpose of contact tracing app is to monitor the movements of citizens, so they suffer from a range of privacy issues. 

\textbf{Mass surveillance.} Since the Government is in charge of the app and of the centralized server, mass surveillance is a plausible threat. G. Avitabile et al.~\cite{rel-prontoc2} enumerate several types of mass surveillance scenarios applied to the DP-3T framework. Infected patients can be easily tracked by an attacker possessing a pervasive infrastructure (e.g., a corporation). The only requirement is a sufficiently large set of devices able to collect \gls{BLE} signals. Infected patients upload their \gls{SK} to the centralized server, and other users are able to calculate \gls{EBID} from them. An attacker can track the movements of infected patients, if they stay nearby the set of devices. Another scenario involves the centralized server colluding with health authorities to map anonymous app users with real identities, since the health authorities themselves perform tests and allow data uploads.

\textbf{Social graph.} In a \emph{centralized} approach, the centralized server stores all of the infected patients \gls{EBID}. S. Vaudenay~\cite{rel-centdecent} argues that the \emph{centralized} server would obtain various lists of pseudonyms unlinkable to real users. Thus, a large monitoring infrastructure and heavy data mining would be required in order to extract meaningful information. The issue has a more limited impact over \emph{decentralized} approaches, since the \gls{ACT} apps store \gls{EBID} locally.

\textbf{Sharing unwanted data.} \gls{ACT} apps may ask for, and share, more data than necessary, and potentially disclose them to third parties. Even though many \gls{ACT} app developers released their app source code, most of them did not release the source code of the centralized server. Thus, there is no guarantee that the disclosed code is the real one. The Indian official \gls{ACT} app, i.e., Aarongya Setu, asks for a lot of sensible information (i.e., name, phone number, age, gender, profession, workplace, recent travels) without proper motivations behind this design choice~\cite{rel-intfree}. It also collects plenty of data using GPS and \gls{BLE}, and the centralized server code is yet to be disclosed. D. Leith et al.~\cite{rel-gaenshare} analyze the data shared through the \gls{GAEN} \gls{API}. On Android devices, the \gls{GAEN} \gls{API} relies on the Google Play Services, which connect to Google servers multiple times a day, sharing quite a lot of sensible data, such as: email address, phone number, IP address (used to retrieve location), phone IMEI, hardware serial number, SIM serial number and Wi-Fi MAC address. 

\subsection{Security Issues of Contact Tracing Apps}
\label{ssec:sec_issues_act_apps}
\gls{BLE} technology allows devices to communicate without consuming a lot of battery power. However, the anonymity constraint, embraced by \gls{ACT} apps to protect the privacy of users, allows any attacker to send false advertisement packets, or tamper with existing ones. Security countermeasures are scarce because of the traditionally low computational power of \gls{BLE} devices.

\textbf{Replay and Relay attacks.} BlueTrace researchers acknowledge the susceptibility of their protocol to replay and relay attacks~\cite{rel-bluetrace}, but do not propose any countermeasure, apart from a human-in-loop methodology. P. Dehaye et al.~\cite{rel-analysswiss} highlight the same issue with replay and relay attacks, and illustrate the idea of tampering with the \gls{RPI} and the \gls{AEM}. Tampering with the transmission power level value would alter the victim's risk score, making it completely unreliable.

\textbf{Denial-of-Service.} M. E. Garbelini et al.~\cite{rel-sweyntooth} uncover how a faulty software implementation of several \gls{BLE} system-on-chip vendors exposes devices to two families of vulnerabilities. Malicious attackers in radio range can manufacture and send specific \gls{BLE} packets to trigger deadlocks, crashes and buffer overflows. Smartphones may incorporate one of the affected system-on-chip, as a part of their Bluetooth module.
\section{Discussion and Conclusion}
\label{sec:discussion}
Contact tracing is used to monitor the spread of SARS-CoV-2, but suffers from scalability issues, since it requires human intervention all the way through. Mobile apps are extremely powerful assets in the scope of \gls{ACT}: they benefit from the popularity of mobile devices among the population, allow for pervasive proximity tracing and gather data in complete autonomy. Many European and Asian countries already released their own national contact tracing app, since each government could choose among the several protocols proposed by different actors: government agencies and commissions (i.e., BlueTrace), researchers (i.e., PACT, ROBERT, DP-3T), or vendors of the leading mobile \gls{OS} (i.e., GAEN). To be successful, \gls{ACT} solutions require consensus from the citizens, since they have to install and use them on a voluntary basis. Thus, adoption rate is the most valuable metric: A. Galanopoulos et al.~\cite{rel-measurement} describe how the current adoption rate varies between 5\% and 20\% for many countries, while the scientific community debates that, to be effective, \gls{ACT} solutions require at least an adoption rate of 60\%. TraceTogether (Singapore) was downloaded by more than one million users, representing the 17\% of Singaporean citizens. Aarongya Setu (India) achieved a total of 160 million downloads, amounting to 12\% of the Indian population. Immuni (Italy) reached an adoption rate of 12.5\% in October 2020, with more than 8 million downloads. COVID Tracker~\cite{ct-covidtracker} (Ireland) is possibly the contact tracing app with the highest adoption rate in Europe, standing at 34\%. It identifies approximately 6 close contacts per single positive case, while the South Korean mass surveillance system identifies slightly more than 10 close contacts per case~\cite{web-silverb}. There are multiple causes for such low adoption rates. User reticence is related to \gls{ACT} unreliability and to the fear of privacy violation. D. Zeinalipour-Yazti et al.~\cite{rel-adoption} tested Swiss, German and Italian contact tracing apps risk score calculation, and argue that \gls{BLE} signal strength has little correlation to physical proximity of app users. If that was the case, the foundations of the whole \gls{ACT} process would fail. Citizens are worried about the Government spying on them, and exploiting the critical situation to start a mass surveillance program. Bluetooth proximity tracing protocols offer better privacy than GPS ones, since they do not share any location data. However, they are still vulnerable to collusion between the Government and the Public Health Authority, and to classic security threats, such as relay, replay and \gls{DoS} attacks. Possible consequences would be the disclosure of sensitive information, surveillance, quarantine enforcement, identity theft and data tampering. 

In this paper, we focused on the resilience to relay attacks of all the \gls{ACT} solutions and, in particular, on GAEN-based apps. We found out that they are vulnerable to such attacks and we were able to develop an attack proof of concept against \emph{Immuni}, the Italian \gls{GAEN}-based app. By designing two malicious apps, we managed to capture the official \emph{Immuni} advertising packets from a victim app with the first malicious app and to relay them through the second malicious app, to a designed victim. We were able to impersonate an official \gls{GAEN}-based app just by re-transmitting the original \gls{BLE} advertising packets. To tackle the vulnerability we found out, we designed \emph{ACTGuard}, a solution that is compliant with current \gls{GAEN} design, while providing a defence against relay attacks and retaining the same privacy features of existing \gls{GAEN}-based apps. \emph{ACTGuard} locally stores the location of each contact between two app users, together with \gls{RPI} of those two users and the time of the contact. However, by saving such information as the result of a hash function, \emph{ACTGuard} fully guarantees the privacy of its users, even when that information is shared with a remote server, functioning as a database for infected user hashes. We implemented a proof of concept of \emph{ACTGuard} by considering \emph{Immuni}, thus releasing the \emph{ImmuniGuard} app. We demonstrated how Immuni is vulnerable to relay attacks and how \emph{ImmuniGuard} can effectively detect a relay attempt by finding inconsistencies between data downloaded from \emph{Immuni} server and from \emph{ImmuniGuard} server.


\balance

\bibliographystyle{IEEEtran}
\bibliography{biblio}

\end{document}